\documentclass[trackchanges,twocolumn]{aastex701}

\usepackage[T1]{fontenc}
\usepackage{CJKutf8}
\usepackage{graphicx}	% Including figure files
\usepackage{amsmath}	% Advanced maths commands
\usepackage{bm}

\usepackage[dvipsnames]{xcolor}
\newcommand{\change}[1]{{#1}} 

\begin{document}

\title{Diffusive Instabilities in Dusty Disks: Linear Growth and Nonlinear Breakdown}

\begin{CJK*}{UTF8}{bsmi}

\author[orcid=0000-0002-4836-1310]{Konstantin Gerbig}
\altaffiliation{Eric and Wendy Schmidt AI in Science Fellow}
\affiliation{Department of Astronomy \& Astrophysics, University of Chicago, Chicago, IL 60637, USA}
\affiliation{Department of Astronomy, Yale University, New Haven, CT 06511, USA}
\email[show]{kgerbig@uchicago.edu}  

\author[orcid=0000-0002-8597-4386]{Min-Kai Lin (林明楷)} 
\affiliation{Institute of Astronomy and Astrophysics, Academia Sinica, Taipei 10617, Taiwan}
\affiliation{Physics Division, National Center for Theoretical Sciences, Taipei 10617, Taiwan}
\email{}


\begin{abstract}
We revisit the diffusive instability in dusty disks that arises when the dust mass diffusivity and/or viscosity decreases sufficiently steeply with increasing dust density. Our updated model includes an incompressible, viscous gas that responds azimuthally and couples to the dust through drag. We show that the basic criterion for diffusion-slope-driven instability remains approximately $\beta_\mathrm{diff}\lesssim -2$ for small dust stopping times, with gas feedback providing only modest quantitative changes for parameters motivated by streaming-instability turbulence. We perform nonlinear numerical calculations and confirm linear growth and mode selection toward the fastest-growing wavenumber. However, for power-law closures $D\propto\Sigma^{\beta_\mathrm{diff}}$ with $\beta_\mathrm{diff}<0$, the nonlinear evolution does not saturate. Instead, steepening gradients amplify the nonlinear dust-pressure term and drive finite-time collapse into increasingly sharp spikes. Motivated by the absence of multidimensional saturation channels in our 1D framework, we test a simple piecewise closure in which the negative diffusion slope operates only over a finite density interval. This modification eliminates blowup and produces peak densities controlled by the imposed saturation scale. Our results support diffusive instabilities as a linear organizing mechanism in dusty turbulence, while highlighting that realistic nonlinear saturation requires additional physics beyond the present closure.
\end{abstract}



\section{Introduction}

Bottom-up planet formation in protoplanetary disks is thought to involve the assembly of a collection of small dust particles into kilometer-sized planetesimals. As coagulational growth becomes inefficient past meter sizes \citep[e.g.,][]{Birnstiel2016}, additional effects must be invoked that bridge the gap from dust to planetesimals. One such mechanism is the gravitational fragmentation of a sufficiently dense patch of the dusty sub-disk. Possible modes of dust concentration that can achieve such densities have thus been subject to intensive research, including turbulent clustering \citep{Chambers2010, Hartlep2020}, secular gravitational instabilities \citep[][]{Ward2000, Tominaga2019}, dust trapping in pressure bumps \citep[][]{Onishi2017} or vortices \citep[][]{Liu2023}, and drag instabilities \citep{Squire2018} such as the streaming instability \citep{Youdin2005}. 

Recently, we described a new, linear instability in dusty disks that operates if mass diffusion, shear viscosity, or both decrease sufficiently rapidly with increasing dust surface mass density \citep[][]{Gerbig_Lin_Lehmann_2024}. The devised model is dust-only and agnostic to the source of diffusion. The permitted modes are thus intrinsic to the dust fluid, and, importantly, not related to dust-gas drag instabilities. Instead, we hypothesized, that the devised model be applied to the nonlinear turbulent quasi-steady state in which the linear streaming instability saturates in, and as such, may be responsible for the near-ubiquitous emergence of azimuthally elongated dust filaments that are seen in numerical simulations \citep[e.g.,][]{Johansen2009, Yang2014, Schreiber2018, Li2018, Abod2019, Gerbig2020, Li2021, Schaefer2024, Lim2025, Li2025}. 

The fast growth rates, the requirements on the diffusion slope that are seemingly met in numerical simulations \citep[][]{Schreiber2018, Gerbig2023}, and the model's hypothesized applicability to turbulence self-generated by streaming instability, warrant further investigation of the diffusive instabilities. 

In this paper, we undertake this task, relax a number of simplifying assumptions that were made in the model in \citet[][]{Gerbig_Lin_Lehmann_2024}, and reassess the parameter space in which diffusive instabilities operate. For instance, in \citet[][]{Gerbig_Lin_Lehmann_2024} the dust fluid was wholly decoupled from the gas -- a good assumption if gas velocity perturbations are negligible -- with the advantage being that the two-dimensional analysis is then reduced to three equations that are analytically tractable.

In the current paper, we do not make this assumption, and instead allow azimuthal velocity perturbations in a viscous gas, which is assumed incompressible. We also include the gas-dust drag feedback terms that explicitly depend on the dust-gas mass ratio, a quantity that did not enter the simplified problem in \citet[][]{Gerbig_Lin_Lehmann_2024}. This dust-gas feedback is dynamically important if it dominates over the gas viscous stress, and the first goal of this paper is to understand the impact of associated gas perturbations for diffusive instabilities. The second goal of this paper is to explore the diffusive instability beyond linear perturbation theory using non-linear, numerical calculations with eigenvalue and random noise initial conditions, for which we employ a custom solver cross-checked with the  \texttt{Dedalus} code \citep[][]{Burns2020}. The focus of our numerical analysis is the modes associated with the parameter space appropriate for streaming instability turbulence, in order to explore the diffusive instability's relevance in planetesimal formation. As the utilized dust pressure prescription does not permit nonlinear saturation, the final goal of the paper is to explore how an alternative, saturating closure relations may reconcile the nonlinear evolution of the diffusive instability with nonlinear streaming instability simulations.

This paper is organized as follows. In Sect.~\ref{sect:model}, we present the updated (nonlinear) model, derive the linearized equations and the diffusive instabilities' dispersion relation. Sect.~\ref{sect:numericaltests} contains our numerical calculations of the nonlinear model, and Sect.~\ref{sect:absenceofsaturation} deals with the lack of nonlinear saturation. We summarize and conclude in Sect.~\ref{sect:discussion}. 

% Dust–gas dynamics driven by the streaming instability with variouspressure gradients
%Stanley A. Baronett

\section{Model}
\label{sect:model}

\subsection{Dust Equations}

Consider an advection-diffusion equation for a dust fluid with density $\rho_\mathrm{d}$ and velocity $\bm{v}$
\begin{align}
    \frac{\partial \rho_\mathrm{d}}{\partial t} + \nabla \cdot \left(\rho_\mathrm{d}\bm{v}\right) = \nabla \cdot \bm{F},
\end{align}
where $\bm{F}$ is some diffusion flux. \change{In a Reynolds-decomposition framework, $\bm{F}$ can be interpreted as arising from correlations between density and velocity fluctuations, i.e. $\bm{F} = \langle \Delta \rho_\mathrm{d}\, \Delta \bm{v} \rangle$.} We define
\begin{align}
    \bm{v}_\mathrm{diff} \equiv - \bm{F}/\rho_\mathrm{d},
\end{align}
such that the continuity equation can be written as
\begin{align}
    \frac{\partial \rho_\mathrm{d}}{\partial t} + \nabla \cdot \left[\rho_\mathrm{d}\left(\bm{v}+\bm{v}_\mathrm{diff}\right)\right] = 0.
\end{align}
Denoting gas velocity as $\bm{u}$, dust stopping time as $t_\mathrm{s}$, dust velocity dispersion as $c_\mathrm{d}$, and dust viscous stress tensor as $T$, the pressurized dust momentum equation \citep[][]{Huang2022} 
\begin{align}
\begin{split}
        \frac{\partial}{\partial t}\left[\rho_\mathrm{d}\left(\bm{v}+\bm{v}_\mathrm{diff}\right)\right] + \nabla \cdot \left[\rho_\mathrm{d}\left(\bm{v}\bm{v} + \bm{v}\bm{v}_\mathrm{diff} +\bm{v}_\mathrm{diff}\bm{v}\right)\right] \\ = \frac{\rho_\mathrm{d}}{t_\mathrm{s}} \left(\bm{u} - \bm{v}\right) - \nabla \left(c_\mathrm{d}^2 \rho_\mathrm{d}\right) + \nabla \cdot T,
\end{split}
\end{align}
can be written as
\begin{align}
\begin{split}
        \frac{\partial (\bm{v} + \bm{v}_\mathrm{diff})}{\partial t} + \left(\bm{v} + \bm{v}_\mathrm{diff}\right)\nabla \cdot \left(\bm{v} + \bm{v}_\mathrm{diff}\right) \\ = \frac{1}{\rho_\mathrm{d}}\nabla (\rho_\mathrm{d} \bm{v}_\mathrm{diff} \bm{v}_\mathrm{diff}) + \frac{\bm{u} - \bm{v}}{t_\mathrm{s}} \\ - \frac{1}{\rho_\mathrm{d}}\nabla \left(c_\mathrm{d}^2 \rho_\mathrm{d}\right) + \frac{1}{\rho_\mathrm{d}}\nabla \cdot T.
\end{split}
\end{align}
We define
\begin{align}
    \overline{\bm{v}} \equiv \bm{v} + \bm{v}_\mathrm{diff}
\end{align}
to write
\begin{align}
\begin{split}
\label{eq:dust_momentum_step3}
        \frac{\partial \overline{\bm{v}}}{\partial t} + \overline{\bm{v}}\nabla \overline{\bm{v}} = \frac{1}{\rho_\mathrm{d}}\nabla \cdot (\rho_\mathrm{d} \bm{v}_\mathrm{diff} \bm{v}_\mathrm{diff}) \\ + \frac{\bm{u} - \overline{\bm{v}}}{t_\mathrm{s}}  + \frac{1}{\rho_\mathrm{d}}\nabla \cdot T + \frac{\bm{v}_\mathrm{diff}}{t_\mathrm{s}} - \frac{1}{\rho_\mathrm{d}}\nabla \left(c_\mathrm{d}^2 \rho_\mathrm{d}\right).
\end{split}
\end{align}
and adjust the viscous stress tensor components accordingly, i.e.
\begin{align}
\begin{split}
    T_{ij} = \nu \rho_\mathrm{d}\Biggr(\frac{\partial (\overline{v}_i - v_{\mathrm{diff}, i})}{\partial x_j}+ \frac{\partial (\overline{v}_j - v_{\mathrm{diff}, j})}{\partial x_i} \\ -\frac{2}{3}\delta_{ij} \nabla \cdot \left(\overline{\bm{v}} - \bm{v}_\mathrm{diff}\right)\Biggr),
\end{split}
\end{align}
with dust viscosity $\nu$. The dust mass diffusion coefficient $D$ is introduced via the closure relations
\begin{align}
\label{eq:dustpressure}
    c_\mathrm{d}^2 &= \frac{D}{t_\mathrm{s}}, \\
    \bm{v}_\mathrm{diff} &= -\frac{D}{\rho_\mathrm{d}} \nabla \rho_\mathrm{d}.
\end{align}
The tie between effective dust pressure and diffusion in Eq.~\eqref{eq:dustpressure} is a key underlying assumption in our formulation and is necessary for linear instability; see Appendix B of \citet[][]{Gerbig_Lin_Lehmann_2024} as well as \citet[][]{Klahr2021}. Further allowing for a power law density dependence in dust diffusion and dust viscosity
\begin{align}
    \beta_\mathrm{D} & \equiv \frac{\partial (D\rho_\mathrm{d})}{\partial \rho_\mathrm
    d} = D (1+\beta_\mathrm{diff}), \\
    \beta_\mathrm{\nu} & \equiv \frac{\partial (\nu\rho_\mathrm{d})}{\partial \rho_\mathrm
    d} = \nu (1+\beta_\mathrm{visc}),
\end{align}
we combine the last two terms in Eq.~\eqref{eq:dust_momentum_step3} into one pressure term
\begin{align}
\begin{split}
    \frac{\partial \overline{\bm{v}}}{\partial t} + \overline{\bm{v}}\nabla \overline{\bm{v}}   = \frac{1}{\rho_\mathrm{d}}\nabla \cdot (\rho_\mathrm{d} \bm{v}_\mathrm{diff} \bm{v}_\mathrm{diff}) \\ + \frac{\bm{u} - \overline{\bm{v}}}{t_\mathrm{s}} + \frac{1}{\rho_\mathrm{d}}\nabla \cdot T - \frac{1}{\rho_\mathrm{d}} \frac{\beta_D+D}{t_\mathrm{s}}\nabla  \rho_\mathrm{d}, 
    \end{split}
\end{align}
or in terms of the dimensionless diffusion slope
\begin{align} 
\begin{split}
        \frac{\partial \overline{\bm{v}}}{\partial t} + \overline{\bm{v}}\nabla \overline{\bm{v}} = \frac{1}{\rho_\mathrm{d}}\nabla\cdot (\rho_\mathrm{d} \bm{v}_\mathrm{diff} \bm{v}_\mathrm{diff}) \\ + \frac{\bm{u} - \overline{\bm{v}}}{t_\mathrm{s}} -\frac{1}{\rho_\mathrm{d}} \frac{D}{t_\mathrm{s}}\left(2+ \beta_\mathrm{diff}\right)\nabla  \rho_\mathrm{d} + \frac{1}{\rho_\mathrm{d}}\nabla \cdot T, 
\end{split}
\end{align}
The viscous stress tensor components read
\begin{align}
\begin{split}
        T_{ij} = \nu \rho_\mathrm{d}\left(\frac{\partial \overline{v}_i}{\partial x_j}+ \frac{\partial \overline{v}_j}{\partial x_i}-\frac{2}{3}\delta_{ij} \nabla \cdot \overline{\bm{v}}  \right) \\
        + \nu \rho_\mathrm{d}\Biggr(\frac{\partial}{\partial x_j} \frac{D}{\rho_\mathrm{d}}\frac{\partial \rho_\mathrm{d}}{\partial x_i}+ \frac{\partial }{\partial x_i}\frac{D}{\rho_\mathrm{d}}\frac{\partial \rho_\mathrm{d}}{\partial x_j}\\-\frac{2}{3}\delta_{ij} \nabla \cdot \left(\frac{D}{\rho_\mathrm{d}}\nabla \rho_\mathrm{d}\right)\Biggr).
\end{split}
\end{align}
The components of the diffusion tensor term read
\begin{align}
\begin{split}
   &\left( \frac{1}{\rho_\mathrm{d}}\nabla\cdot (\rho_\mathrm{d} \bm{v}_\mathrm{diff} \bm{v}_\mathrm{diff})\right)_i \\ &= - \frac{1}{\rho_\mathrm{d}}\sum_j \frac{\partial}{\partial x_j}\left[\frac{D^2}{\rho_\mathrm{d}}\left(\frac{\partial \rho_\mathrm{d}}{\partial x_i}\right)\left(\frac{\partial \rho_\mathrm{d}}{\partial x_j}\right)\right].
\end{split}
\end{align}

\subsection{Shearing-Box Equations}

The shearing box is a co-rotating, cartesian ($x,y,z$) coordinate frame at distance $r_0$ away from the star. Under the assumption that $x,y \ll r_0$, the Keplerian shear can be linearized. Velocities are measured relative to the background shear motion, for a forward pointing $\hat{\bm{y}}$ given by $-q\Omega x$. We define
\begin{align}
    \bm{u}^* &= \bm{u} + q \Omega x \hat{\bm{y}}, \\
    \overline{\bm{v}}^* &= \overline{\bm{v}} + q \Omega x \hat{\bm{y}}.
\end{align}
Adding stellar radial gravity to the momentum equation, yields
\begin{align} 
\begin{split}
        \frac{\partial \overline{\bm{v}}^*}{\partial t} - q\Omega x \frac{\partial \overline{\bm{v}}^*}{\partial y}+ \overline{\bm{v}}^*\nabla \overline{\bm{v}}^* = 2\Omega\overline{v}_y^*\hat{\bm{x}}  \\ - \left(2-q\right)\Omega \overline{v}_x^* \hat{\bm{y}} + \frac{1}{\rho_\mathrm{d}}\nabla\cdot (\rho_\mathrm{d} \bm{v}_\mathrm{diff} \bm{v}_\mathrm{diff})  \\ + \frac{\bm{u}^* - \overline{\bm{v}}^*}{t_\mathrm{s}} + \frac{\bm{v}_\mathrm{diff}}{t_\mathrm{s}} - \frac{1}{\rho_\mathrm{d}}\nabla \left(c_\mathrm{d}^2 \rho_\mathrm{d}\right) + \frac{1}{\rho_\mathrm{d}}\nabla \cdot T, 
\end{split}
\end{align}
Plugging in the closure relations, we get
\begin{align} 
\begin{split}
        \frac{\partial \overline{\bm{v}}^*}{\partial t} - q\Omega x \frac{\partial \overline{\bm{v}}^*}{\partial y}+ \overline{\bm{v}}^*\nabla \overline{\bm{v}}^*  \\ - 2\Omega\overline{v}_y^*\hat{\bm{x}} 
        + \left(2-q\right)\Omega \overline{v}_x^*\hat{\bm{y}}  \\ =   \frac{1}{\rho_\mathrm{d}}\nabla\cdot (\rho_\mathrm{d} \bm{v}_\mathrm{diff} \bm{v}_\mathrm{diff})  + \frac{\bm{u}^* - \overline{\bm{v}}^*}{t_\mathrm{s}} + \frac{1}{\rho_\mathrm{d}}\nabla \cdot T \\ - \left(\frac{2+\beta_\mathrm{diff}}{t_\mathrm{s}}\right)\frac{D}{\rho_\mathrm{d}}\nabla  \rho_\mathrm{d}. 
\end{split}
\end{align}
Next, we assume \change{axial and vertical} symmetry such that $\partial/\partial y = \partial/\partial z = 0$. 
\change{Since we neglect vertical gravity and therefore describe a vertically homogeneous shearing box, we may equivalently adopt a vertically integrated description and replace the volume density $\rho_\mathrm{d}$ by the surface density $\Sigma_\mathrm{d} \equiv \int \rho_\mathrm{d}\,\mathrm{d}z$.}
The component-wise momentum equations are
\begin{align}   
    \begin{split}
    \label{eq:nonlinear_dust_vx}
        \frac{\partial \overline{v}^*_x}{\partial t} + \overline{v}^*_x \frac{\partial \overline{v}^*_x}{\partial x} - 2\Omega \overline{v}^*_y = -\frac{1}{\Sigma_\mathrm{d}}\frac{\partial}{\partial x}\left[\frac{D^2}{\Sigma_\mathrm{d}}\left(\frac{\partial \Sigma_\mathrm{d}}{\partial x}\right)^2\right] \\ 
        + \frac{{u}^*_x- \overline{v}^*_x}{t_\mathrm{s}} - \frac{1}{\Sigma_\mathrm{d}}(2+\beta_\mathrm{diff})\frac{D}{t_\mathrm{s}}\frac{\partial \Sigma_\mathrm{d}}{\partial x} \\ + \frac{4}{3}\frac{1}{\Sigma_\mathrm{d}}\frac{\partial}{\partial x} \left[\nu \Sigma_\mathrm{d}\frac{\partial}{\partial x}\left(\overline{v}^*_x + \frac{D}{\Sigma_\mathrm{d}}\frac{\partial \Sigma_\mathrm{d}}{\partial x}\right)\right].
    \end{split}
\end{align}
\begin{align}
    \label{eq:nonlinear_dust_vy}
    \begin{split}
    \frac{\partial \overline{v}^*_y}{\partial t} +  \overline{v}^*_x \frac{\partial  \overline{v}^*_y}{\partial x} + \left(2-q\right)\Omega \overline{v}^*_x \\
    =  \frac{ {u}^*_y -  \overline{v}^*_y}{t_\mathrm{s}} + \frac{1}{\Sigma_\mathrm{d}}\frac{\partial}{\partial x}\left[\nu\Sigma_\mathrm{d}\left( \frac{\partial \overline{v}^*_y}{\partial x}-q\Omega\right)\right].
    \end{split}
\end{align}
The dust equations are completed with the axisymmetric continuity equation
\begin{align}
\label{eq:nonlinear_cont}
    \frac{\partial \Sigma_\mathrm{d}}{\partial t} + \frac{\partial}{\partial x}\left(\Sigma_\mathrm{d}\overline{v}^*_x\right) = 0.
\end{align}
For closure of the system, we must also treat the gas velocity $\bm{u}^*$. In \citet[][]{Gerbig_Lin_Lehmann_2024}, we considered a static gas with $\bm{u}^* = 0$, which decouples the two fluids and simplifies the system.  Here, we generalize and include an incompressible gas with $\nabla \cdot \bm{u}^* = 0$. For an axisymmetric system, incompressibility implies vanishing radial velocity gradients and thus $u_x^* = 0$, which decouples gas continuity and radial momentum equations from the problem. Closure only requires the azimuthal momentum equation, which under axisymmetry reads
\begin{align}
    \label{eq:nonlinear_gas_ymom}
    \frac{\partial u_y^*}{\partial t} = \frac{\varepsilon}{t_\mathrm{s}}\left(\overline{v}_y^* - {u}_y^*\right) + \nu_\mathrm{g}\frac{\partial^2 u_y^*}{\partial x^2},
\end{align}
 where we introduced (assumed-to be constant) gas viscosity $\nu_\mathrm{g}$, and dust-to-gas ratio $\varepsilon \equiv \rho_\mathrm{d}/\rho_\mathrm{g}$.

Eqs.~\eqref{eq:nonlinear_dust_vx}, \eqref{eq:nonlinear_dust_vy}, \eqref{eq:nonlinear_cont}, and \eqref{eq:nonlinear_gas_ymom} constitute the nonlinear model in this paper. In the following, we will drop both superscripts and overlines of the velocities.

\subsection{Linearized Equations}
\label{sect:linearized_equations}

\change{Unlike many classical diffusive instabilities, such as double-diffusive \citep[e.g.,][]{Garaud2018} or reaction–diffusion instabilities \citep[][]{Turing1952}, this paper's diffusive instability does not rely on the presence of any background gradients or stratification. The background state is strictly homogeneous, and the instability is instead driven by the density-dependence of the transport coefficients themselves. In this sense, the present instability is a constitutive instability: infinitesimal fluctuations grow because regions of enhanced density experience a reduced effective diffusion, which further suppresses smoothing and leads to positive feedback.} We perturb the system around this constant background, i.e. $\Sigma_\mathrm{d} = \Sigma_\mathrm{d,0} + \Sigma_\mathrm{d}^\prime$, $D = D_0 + D^\prime$, $\nu = \nu_0 + \nu^\prime$, $\bm{v} = \bm{v}^\prime$, $\bm{u} = \bm{u}^\prime$, with $\Sigma_\mathrm{d,0} = \mathrm{const.}$, $D_0 = \mathrm{const.}$, $\nu_0 = \mathrm{const.}$, and linearize in perturbed quantities, such that the dust momentum equations become
\begin{align}
    \begin{split}
        \frac{\partial v_x^\prime}{\partial t} = 2\Omega v_y^\prime + \frac{u_x^\prime - v_x^\prime}{t_\mathrm{s}} - \frac{1}{\Sigma_\mathrm{d,0}}(2+\beta_\mathrm{diff})\frac{D_0}{t_\mathrm{s}}\frac{\partial \Sigma_\mathrm{d}^\prime}{\partial x} \\ + \frac{4}{3}\nu_0 \frac{\partial^2}{\partial x^2}\left(v_x^\prime + \frac{D_0}{\Sigma_\mathrm{d,0}}\frac{\partial \Sigma_\mathrm{d}^\prime}{\partial x}\right),
    \end{split}
\end{align}
\begin{align}
\begin{split}
    \frac{\partial v_y^\prime}{\partial t} = - (2-q)\Omega v_x^\prime  + \frac{u_y^\prime - v_y^\prime}{t_\mathrm{s}} \\ +\nu_0 \frac{\partial^2 v_y^\prime}{\partial x^2} - \frac{q\Omega \beta_\nu}{\Sigma_\mathrm{d,0}} \frac{\partial \Sigma_\mathrm{d}^\prime}{\partial x}.
\end{split}
\end{align}
This formalism \change{differs from} \citet[][]{Gerbig_Lin_Lehmann_2024} physically only in that now we account for $\partial (\rho_\mathrm{d}\bm{v}_\mathrm{diff})/\partial t$. In practice, the differences are that (1) velocities are defined relative to the diffusion velocity $\bm{v}_\mathrm{diff}$, leading to (2) the continuity equation not containing a diffusion term, (3) an increased pressure term (`reference frame' pressure in addition to the intrinsic $c_\mathrm{d}^2$ pressure), (4) the lack of diffusion flux terms, and (5) a modified viscous stress due to the diffusion velocity. The $y$-momentum equation looks identical, since axisymmetry has $v_{\mathrm{diff},y} = 0$.

We introduce Fourier modes $f^\prime \propto \hat{f} \exp(-ikx + nt)$, with growth rate $\Re(n)$ and oscillation frequency $\Im(n)$. We get
\begin{align}
    n\frac{\hat{\Sigma}_\mathrm{d}}{\Sigma_\mathrm{d,0}} = & ik \hat{v}_x, \\
\begin{split}
    n \hat{v}_x = & 2\Omega \hat{v}_y - \frac{\hat{v}_x}{t_\mathrm{s}} + \frac{ik}{t_\mathrm{s}}\left(2+\beta_\mathrm{diff}\right)D_0\frac{\hat{\Sigma}_\mathrm{d}}{\Sigma_\mathrm{d,0}} \\ & -\frac{4}{3}\nu_0 k^2 \hat{v}_x  +\frac{4}{3}ik^3\nu_0 D_0\frac{\hat{\Sigma}_\mathrm{d}}{\Sigma_\mathrm{d,0}},
\end{split}
\\
    \begin{split}
        n\hat{v}_y = & -\left(2-q\right)\Omega \hat{v}_x   + \frac{\hat{u}_y-\hat{v}_y}{t_\mathrm{s}}  \\ &- \nu_0 k^2 \hat{v}_y + ikq\nu_0\Omega\left(1+\beta_\mathrm{visc}\right)\frac{\hat{\Sigma}_\mathrm{d}}{\Sigma_\mathrm{d,0}},
    \end{split}
\\
    \label{eq:fourier_mode_gas_y_momentum}
    n\hat{u}_y = & - \frac{\varepsilon}{t_\mathrm{s}}\left(\hat{u}_y - \hat{v}_y\right) - \nu_\mathrm{g}k^2\hat{u}_y.
\end{align}

\subsection{Dispersion Relation}

This system is solved by a fourth-order dispersion relation
\begin{align}
\label{eq:fourth_order_disprel}
    n^4 + a_3 n^3 + a_2 n^2 + a_1 n + a_0 = 0.
\end{align}
Defining the Schmidt numbers for dust and gas as 
\begin{align}
    \mathrm{Sc} \equiv \frac{\nu_0}{D_0}, \
    \mathrm{Sc}_\mathrm{g} \equiv \frac{\nu_\mathrm{g}}{D_0},
\end{align}
respectively, and plugging in $q=3/2$, the coefficients are
\begin{align}
\begin{split}
    a_0 = & \frac{16 \mathrm{Sc}^2\mathrm{Sc}_\mathrm{g}D^4 k^8}{9} + \frac{4\mathrm{Sc}D^3 k^6}{3t_\mathrm{s}}\left[\varepsilon + \mathrm{Sc}_\mathrm{g}\left(\frac{10}{3} + \beta_\mathrm{diff}\right)\right] \\
    & + D^2 k^4 \Bigg\{\mathrm{Sc}_\mathrm{g}\left[4\mathrm{Sc}\Omega^2\left(1+\beta_\mathrm{visc}\right) + \frac{4}{3t_\mathrm{s}^2}\left(2+\beta_\mathrm{diff}\right)\right] \\ & + \frac{\mathrm{Sc}\varepsilon}{t_\mathrm{s}^2}\left(2+\beta_\mathrm{diff}\right)\Bigg\}
     + \frac{3\varepsilon\mathrm{Sc}\Omega^2 D k^2}{t_\mathrm{s}}\left(1+\beta_\mathrm{visc}\right),
\end{split}
\end{align}
\begin{align}
\begin{split}
    a_1 = &\frac{4\mathrm{Sc}D^3k^6}{3}\left[\mathrm{Sc} + \frac{4}{3}\mathrm{Sc}_\mathrm{g}\left(1+\mathrm{Sc}\right)\right] \\ & + \frac{2D^2k^4}{3t_\mathrm{s}}\Bigg\{2\mathrm{Sc}_\mathrm{g}\left(2 + \frac{7}{3}\mathrm{Sc}+3\beta_\mathrm{diff}\right) \\ & + \mathrm{Sc}\left[5+2\varepsilon\left(1+\mathrm{Sc} \right)+ 3\beta_\mathrm{diff}\right]\Bigg\} \\ & + Dk^2 \Bigg\{\frac{4}{3}\mathrm{Sc}_\mathrm{g}\left(\Omega^2 + \frac{1}{t_\mathrm{s}^2}\right)  \\
    & +\mathrm{Sc}\left[\frac{\varepsilon}{t_\mathrm{s}^2} + 3\Omega^2\left(1+\beta_\mathrm{visc}\right)\right] \\
    & + \frac{1+\varepsilon}{t_\mathrm{s}^2}\left(2+\beta_\mathrm{diff}\right)\Bigg\} + \frac{\varepsilon \Omega^2}{t_\mathrm{s}},
\end{split}
\end{align}
\begin{align}
\begin{split}
    a_2 = & \frac{\mathrm{Sc}D^2k^4}{3}\left[\frac{28}{3}\mathrm{Sc}_\mathrm{g}+ 4\left(1+ \mathrm{Sc}\right)\right] \\
    & + \frac{Dk^2}{t_\mathrm{s}}\left[\frac{8}{3}\mathrm{Sc}_\mathrm{g}+\frac{7}{3}\mathrm{Sc}\left(1+\varepsilon\right) + 2 + \beta_\mathrm{diff}\right]  \\
    & + \frac{1+\varepsilon}{t_\mathrm{s}^2} + \Omega^2,
\end{split}
\end{align}
\begin{align}
\begin{split}
    a_3 = & \frac{D k^2}{3}\left(7\mathrm{Sc} + 4 \mathrm{Sc}_\mathrm{g}\right) + \frac{2+\varepsilon}{t_\mathrm{s}}.
\end{split}
\end{align}
A sufficient criterion for an unstable static mode is $a_0 < 0$. As all quantities are positive with the exception of $\beta_\mathrm{diff}$ and $\beta_\mathrm{visc}$, it is immediately evident that $\beta_\mathrm{diff} < -2$ and $\beta_\mathrm{visc} < -1$ are the approximate criteria for diffusive instability driven by the diffusion slope and viscosity slope respectively. This result of the Galilean-variant, dust-only model in \citet{Gerbig_Lin_Lehmann_2024} is therefore not modified. \change{We further note that, in the limit $D_0 \rightarrow 0$, the destabilizing term vanishes and the system becomes linearly stable, as expected. In contrast, for large $D_0$ the instability formally survives, but shifts to progressively smaller wavenumbers. For $D_0 \gtrsim 1$, the unstable modes occur on scales larger than those physically meaningful within the assumptions of the adopted framework.}

\subsection{Growth Rates and Most Unstable Mode}

\begin{figure*}[t]
    \centering
    \includegraphics[width=\linewidth]{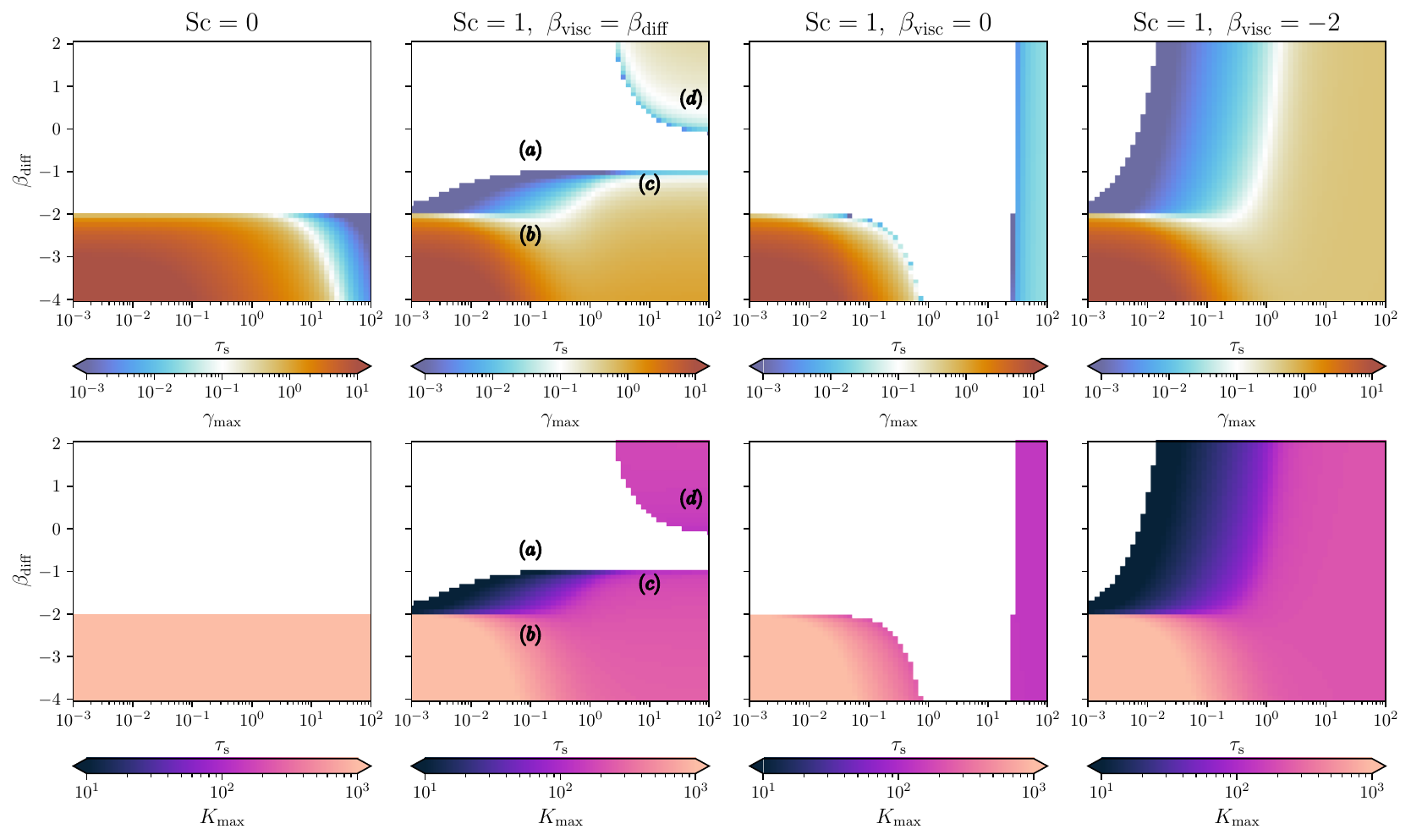}
    \caption{Growth rates $\gamma_\mathrm{max}$ (top row) for the fastest growing modes $K_\mathrm{max}$ (bottom row) of the diffusive instability in $\beta_\mathrm{diff}$ - $\tau_\mathrm{s}$ - space for different values of $\mathrm{Sc}$ and $\mathrm{Sc}_\mathrm{g}$ in different columns. Parameters kept constant between the panels are $\delta = 10^{-5}$, $\alpha_\mathrm{g} = 10^{-3}$, $\varepsilon = 1$. All three `types' of instabilities discussed in \citet[][]{Gerbig_Lin_Lehmann_2024} are present in the depicted parameter space: the diffusive instability driven by the diffusion slope is permitted for sufficiently negative $\beta_\mathrm{diff}$ and sufficiently small $\tau_\mathrm{s}$ (bottom left region of all panels), the diffusive instability driven by the viscosity slope is present for a sufficiently negative viscosity slope $\beta_\mathrm{visc}$ and large $\tau_\mathrm{s}$ (bottom right of second column, and right side of fourth column), and the diffusive overstability driven by the viscosity slope is present for sufficiently positive $\beta_\mathrm{visc}$ and large $\tau_\mathrm{s}$ (top right of second column, and right side of third column). Letters in the second column correspond to setups in the numerical experiments in Sect.~\ref{sect:numericaltests}. }
    \label{fig:instability_map}
\end{figure*}

Figure~\ref{fig:instability_map} shows the growth rates for the fastest growing mode, obtained by finding the roots of Eq.~\eqref{eq:fourth_order_disprel}. We hereby utilize the following dimensionless quantities
\begin{align}
\label{eq:dimless1}
   & \tau_\mathrm{s} \equiv t_\mathrm{s}\Omega, \  \sigma \equiv n/\Omega, \ K \equiv kH, \\
   \label{eq:dimless2}
    & \delta_0 \equiv D_0/(c_\mathrm{s}H), \ \alpha_0 \equiv \nu_0/(c_\mathrm{s}H), \ \alpha_\mathrm{g} \equiv \nu_\mathrm{g}/(c_\mathrm{s}H).
\end{align}

The left-most column of Fig.~\ref{fig:instability_map} depicts the $\mathrm{Sc} = 0$ case. The relevant criterion can be easily obtained by plugging in 
\begin{align}
    \xi \equiv \delta_0 K^2
\end{align}
and dropping all terms $\mathcal{O}(\xi^2)$ or higher. The constant term coefficient drops, and the dimensionless version of the dispersion relation Eq.~\eqref{eq:fourth_order_disprel} becomes third order, i.e. 
\begin{align}
\begin{split}
        \sigma^3 + &\left(\frac{4}{3}\mathrm{Sc}_\mathrm{g}\xi + \frac{2+\varepsilon}{\tau_\mathrm{s}}\right)\sigma^2 \\ & + \left[\left(\frac{8}{3}\mathrm{Sc}_\mathrm{g}+2+\beta_\mathrm{diff}\right)\frac{\xi}{\tau_\mathrm{s}}+ \frac{1+\varepsilon}{\tau_\mathrm{s}^2} + 1\right]\sigma \\
        & + \left[\frac{4}{3}\mathrm{Sc}_\mathrm{g}\left(1+\frac{1}{\tau_\mathrm{s}^2}\right)+\frac{1+\varepsilon}{\tau_\mathrm{s}^2}\left(2+\beta_\mathrm{diff}\right)\right]\xi \\ & + \frac{\varepsilon}{\tau_\mathrm{s}} = 0.
\end{split}
\end{align}
For $|\sigma| \ll 1$, we find an eigenfrequency of
\begin{align}
\label{eq:sigma_approx}
    \sigma \approx  -\frac{4\xi\mathrm{Sc}_\mathrm{g}\left(1+\tau_\mathrm{s}^2\right)/3 + \varepsilon \tau_\mathrm{s} + \left(1+\varepsilon\right)\left(2+\beta_\mathrm{diff}\right)\xi}{\tau_\mathrm{s}^2 +\xi\tau_\mathrm{s}\left(8\mathrm{Sc}_\mathrm{g}/3 + 2 +\beta_\mathrm{diff}\right) + \varepsilon + 1}.
\end{align}
We see that for $\tau_\mathrm{s}\ll 1$,
\begin{align}
    \gamma \equiv \Re{(\sigma)} \approx - \left[(2+\beta_\mathrm{diff}) + \frac{4\mathrm{Sc}_\mathrm{g}}{3(1+\varepsilon)}\right]\xi,
\end{align}
readily confirming that $\beta_\mathrm{diff} \lesssim -2$ leads to instability as seen in Fig.~\ref{fig:instability_map}. The inclusion of dust-gas feedback formally stabilizes. However, if the gas is more \change{viscous} than the dust is diffusive, i.e. $\nu_\mathrm{g} \gg D_0$ (which seems plausible), and/or if dust-to-gas ratio is high, i.e. $\varepsilon \gg 1$, this effect becomes negligible and we recover the dust-only model in \citet{Gerbig_Lin_Lehmann_2024}. This is very much expected given the form linearized gas equation in Eq.~\eqref{eq:fourier_mode_gas_y_momentum}. 

If dust viscosity is included $\mathrm{Sc} > 0$, the model is regularized \citep[as opposed to growing on arbitrarily small scales, see first column of Fig.~\ref{fig:instability_map} and][]{Gerbig_Lin_Lehmann_2024}. Moreover, a dust viscosity slope can also drive instability (columns two and four of Fig.~\ref{fig:instability_map}) or overstability (columns two and three of Fig.~\ref{fig:instability_map}). While the viscosity slope formally also allows instability if $\beta_\mathrm{diff} > -2 $ as long as $\beta_\mathrm{visc}$ is sufficiently negative, growing modes in this regime possess very large wavenumbers with uninterestingly small growth rates if stopping times are small $\tau_\mathrm{s} \lesssim 1$. Only for large stopping times $\tau_\mathrm{s} \gtrsim 1$, the viscosity slope can drive instability and overstability with fast growth rates. While we still include this limit in Fig.~\ref{fig:instability_map}, we do not discuss it further as the fluid description of the dust breaks down \citep[e.g.][]{Garaud2004} and instead refer to Sect.~5.5 in \citet[][]{Gerbig_Lin_Lehmann_2024} for further discussion of the prospects for diffusive instability at large stopping times. 

For the remainder of the paper, which entails a numerical investigation of the model, we focus on a parameter region where $\mathrm{Sc} = 1$, such that the model is regularized, and set $\beta_\mathrm{visc} = \beta_\mathrm{diff}$. This corresponds to the second column of Fig.~\ref{fig:instability_map}.

\section{Numerical Tests}

\label{sect:numericaltests}

\begin{figure*}[t]
    \includegraphics[width=\linewidth]{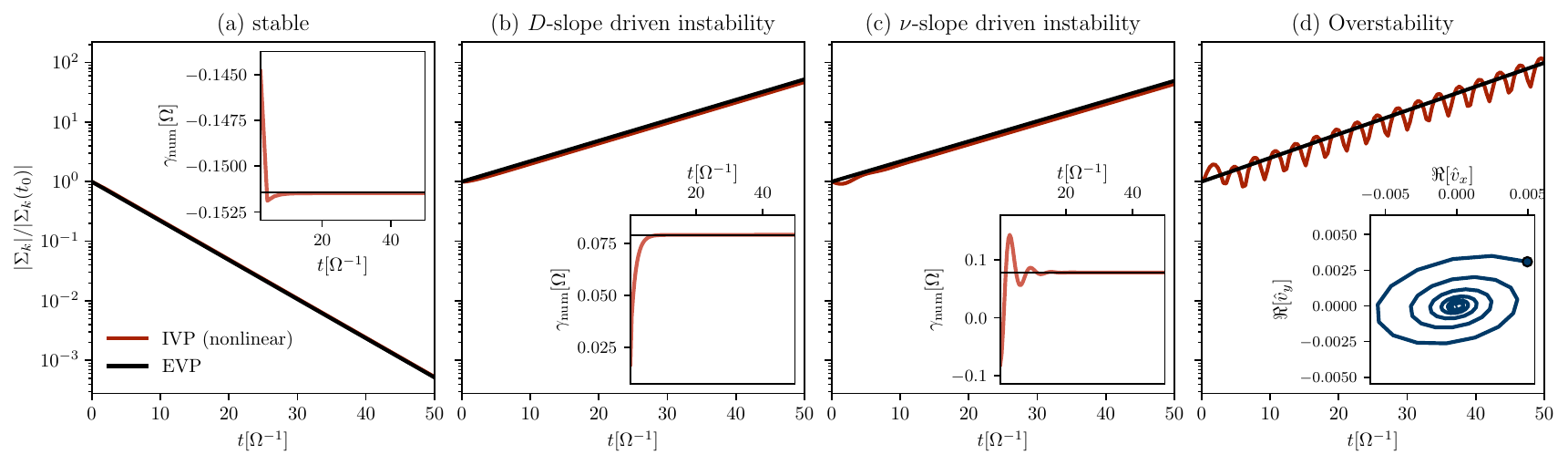}
    \caption{Linear-nonlinear consistency test for different branches of the diffusive instability. An exact linear eigenmode solution for $K=100$ is evolved forward in time using the pseudo-spectral IMEX solver. Panels show the time evolution of the amplitude of the dominant Fourier component of the normalized surface density perturbation (red), compared to the prediction from the linear eigenvalue problem (black). Panels (a) - (d) correspond, respectively, to: (a) a stable, exponentially damped mode ($\tau_\mathrm{s} = 0.1, \beta_\mathrm{visc} = \beta_\mathrm{diff} = -0.5$); (b) the diffusion-slope-driven instability ($\tau_\mathrm{s} = 0.1, \beta_\mathrm{visc} = \beta_\mathrm{diff} = -2.5$); (c) the (dust) viscosity-slope-driven instability ($\tau_\mathrm{s} = 10, \beta_\mathrm{visc} = \beta_\mathrm{diff} = -1.3$); and (d) the overstable mode driven by the (dust) viscosity-slope ($\tau_\mathrm{s} = 50, \beta_\mathrm{visc} = \beta_\mathrm{diff} = 0.7$). The setups are marked in the instability map in Fig.~\ref{fig:instability_map}. The insets in panels (a)-(c) show how the instantaneous numerical growth rate converge to the analytic eigenvalue. The inset in panel (d) shows the spiralling trajectory of the overstable mode in the $(\Re(\hat{v}_x), \Re(\hat{v}_y))$ phase space.}
    \label{fig:linearevolution_branches}
\end{figure*}

We employ a one-dimensional pseudo-spectral solver using an implicit-explicit (IMEX) time-integration strategy to perform a series of numerical tests and evolve the nonlinear, axisymmetric model in Eqs.~\eqref{eq:nonlinear_dust_vx}, \eqref{eq:nonlinear_dust_vy}, \eqref{eq:nonlinear_cont}, and \eqref{eq:nonlinear_gas_ymom} forward in time. There, the linear, constant-coefficient part of the viscous operator, i.e. the terms proportional to $\nu_0\partial_x^2 v_x, \nu_0\partial_x^2 v_y$ and $\nu_\mathrm{g}\partial_x^2 u_y$, is treated implicitly, whereas all remaining terms, including the variable-coefficient corrections to (dust) viscosity and all dust-diffusion, advection, drag and pressure terms are evaluated explicitly. This approach removes the diffusive stability constraint, where the timestep is restricted to $\mathcal{O}(\Delta x^2/\nu_\mathrm{0})$, while retaining a simple explicit treatment of the nonlinear dynamics. 

As an external benchmark, we also solve the same system using the spectral code \texttt{Dedalus} \citep[][]{Burns2020}. For both solvers and across all setups, we choose periodic boundary conditions, a standard dealiasing factor of $2/3$, and a second order Runge-Kutta timestepper. In \texttt{Dedalus}, the hyperbolic and drag terms are handled explicitly while all diffusive and viscous terms are evaluated implicitly, so the underlying scheme is analogous to our native integrator but implemented within a general-purpose PDE framework. The drawback of this generality is a slowdown of about a factor of 10 relative to our specialized pseudo-spectral solver. Both the native solver and the \texttt{Dedalus} backend \change{are publicly available as open-source software \citep{Gerbig_diffinst_zenodo}.}

In the following, we verify the linear growth rates for the branches (Sect.~\ref{sect:lin_validation}), compare the gas-inclusive model to the dust-only counterpart (Sect.~\ref{sect:comp_to_dust_only}), investigate the nonlinear growth and eventual blow-up of modes across several resolutions (Sect.~\ref{sect:nonlinaerblowup}), and evolve a noise initial condition to demonstrate that the dominant Fourier mode in the numerical calculations converges toward the fastest-growing eigenmode predicted by linear theory (Sect.~\ref{sect:randomnoise}). \change{Hereby, we employ the dimensionless code units as defined in Eqs.~\eqref{eq:dimless1} and \eqref{eq:dimless2}.}

\subsection{Linear Validation of Eigenmodes}

\label{sect:lin_validation}

To validate the method, we seed clean eigenmode solutions to the linear problem in Eq.~\eqref{eq:fourth_order_disprel} as initial conditions and compare their (nonlinear) numerical evolution to the linear theory. Hereby, the radial domain size is set to $L_x = 4\pi/K$ to fit two wavelengths at a spectral resolution of $N_x = 128$. We choose $K= 100$. Tests are performed for all branches, i.e. sets of parameters that permit diffusive instability driven by diffusion slope ($\tau_\mathrm{s} = 0.1, \beta_\mathrm{visc} = \beta_\mathrm{diff} = -2.5$), diffusive instability driven by viscosity slope ($\tau_\mathrm{s} = 10, \beta_\mathrm{visc} = \beta_\mathrm{diff} = -1.3$), diffusive overstability driven by viscosity slope ($\tau_\mathrm{s} = 50, \beta_\mathrm{visc} = \beta_\mathrm{diff} = 0.7$), as well as an exponentially damped mode ($\tau_\mathrm{s} = 0.1, \beta_\mathrm{visc} = \beta_\mathrm{diff} = -0.5$). For all solutions, $\delta = \alpha = 10^{-5}$, $\alpha_\mathrm{g} = 10^{-4}$, and $\varepsilon = 1$ is fixed. Fig.~\ref{fig:linearevolution_branches} shows the evolution of the surface density amplitude for the four setups. The inset in panels a)-c) show how the growth rate matches the linear theory prediction. The inset in panel d) shows the mode structure of the oscillatory growing mode. For all branches, we find excellent agreement with the linear-theory prediction of the growth rate. 

For the remaining numerical tests we focus on the diffusive instability driven by the diffusion slope. It is the physically most relevant unstable branch permitted by our model, as it hinges neither on the existence of large stopping time particles (and the matter of the model becoming inappropriate in this regime, see Sect.5.5 in \citet[][]{Gerbig_Lin_Lehmann_2024}), nor a dust viscosity slope which has hitherto not been measured in the setups \change{that are relevant for planetesimal formation}. The diffusive instability driven by the diffusion slope, generally speaking, requires well to marginally coupled particles, i.e. $\tau_\mathrm{s} \lesssim 1$, which applies to most (if not all) dust grains and pebbles in disks; as well as a sufficiently negative diffusion slope $\beta_\mathrm{diff}\lesssim-2$. Measurements of the diffusion slope in simulations \citep[][]{Schreiber2018, Gerbig2023} suggest that this condition may be met. 

\begin{figure*}[t]
    \includegraphics[width=\linewidth]{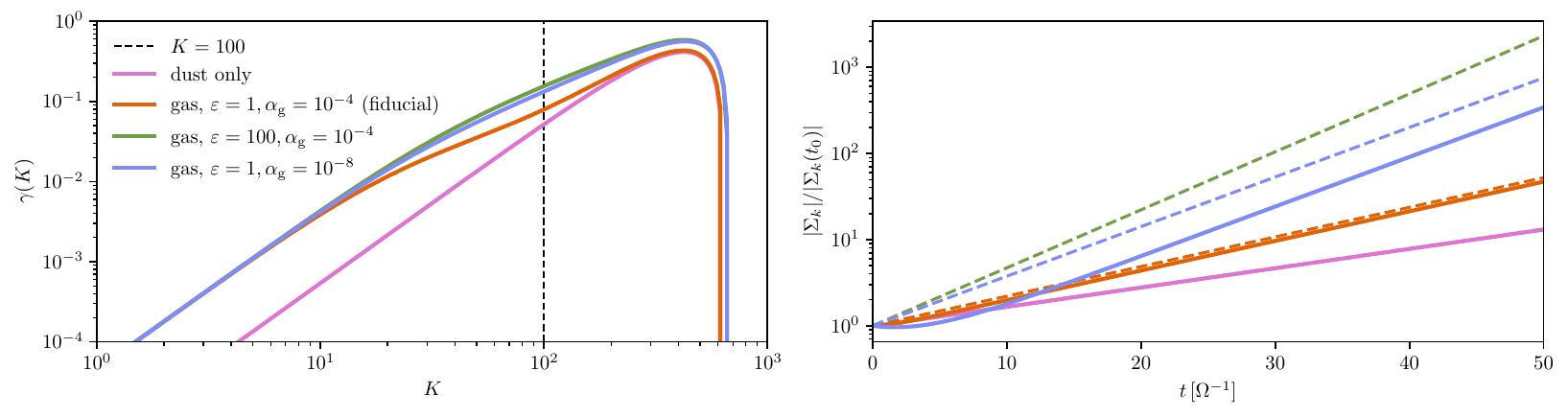}
    \caption{Comparison between dust-only models, testing different values for dust-to-gas ratio $\varepsilon$ and gas viscosity $\alpha_\mathrm{g}$. All setups share $\delta = \alpha = 10^{-5}$, $\tau_\mathrm{s} = 0.1, \beta_\mathrm{visc} = \beta_\mathrm{diff} = -2.5$ such that a $K=100$ mode is subject to diffusive instability driven by diffusion slope. Left: Linear growth rates from the eigenvalue problem. Right: Evolution of a $K=100$ eigenmode with amplitudes that remain in the linear regime.}
    \label{fig:linearevolution_dustvgas}
\end{figure*}

\subsection{Comparison to the Dust-only Model}

\label{sect:comp_to_dust_only}

To quantify the effect of including the incompressible gas component, we test a range of dust-to-gas ratios $\varepsilon$ and gas viscosities $\alpha_\mathrm{g}$ and compare to a dust-only model. All setups share the same `dust parameters', i.e. $\delta = \alpha = 10^{-5}$, $\tau_\mathrm{s} = 0.1, \beta_\mathrm{visc} = \beta_\mathrm{diff} = -2.5$. The left panel of Fig.~\ref{fig:linearevolution_dustvgas} shows the corresponding dispersion relations. Note that the `dust-only' model in Fig.~\ref{fig:linearevolution_dustvgas} is not exactly the model in \citet[][]{Gerbig_Lin_Lehmann_2024} where we did not account for $\partial(\rho_\mathrm{d}\bm{v}_\mathrm{diff})/\partial t$ (see Sect.~\ref{sect:linearized_equations}). Including the gas generally increases growth rates. This is particular pronounced if dust-to-gas ratios are high and/or gas viscosities very low. For the fiducial set up with $\varepsilon = 1$ and $\alpha_\mathrm{g} = 10^{-4}$ the boost is only appreciable at intermediate-to-high wavenumbers. 

The gas is incompressible and only responds azimuthally, so it does not add inertia. Instead, $u_y$ responds  to the $v_y$ perturbation with a finite time lag that is controlled by $t_\mathrm{s}$, $\varepsilon$ and $\alpha_\mathrm{g}$, see Eq.~\eqref{eq:fourier_mode_gas_y_momentum}. This delayed response then feeds back into the $y$-momentum equation with the correct phase to reduce the effective $(u_y-v_y)/t_\mathrm{s}$ drag acting on the dust, which amplifies the motions that drive the instability. This is in contrast to the dust-only case where the dust is simply dragged against a rigid background, i.e. the same term reduces to $-v_y/t_\mathrm{s}$. Still, for typical parameters, i.e. $\varepsilon \gtrsim 1, 10^{-2} \gtrsim \alpha_\mathrm{g} \gtrsim 10^{-6}$, the resulting amplification is modest at best, and the existence of the instability depends far more on the steepness of the diffusion (and dust viscosity) slope than on whether gas is allowed to co-oscillate.

\subsection{Nonlinear Blowup}
\label{sect:nonlinaerblowup}

\begin{figure*}
    \includegraphics[width=\linewidth]{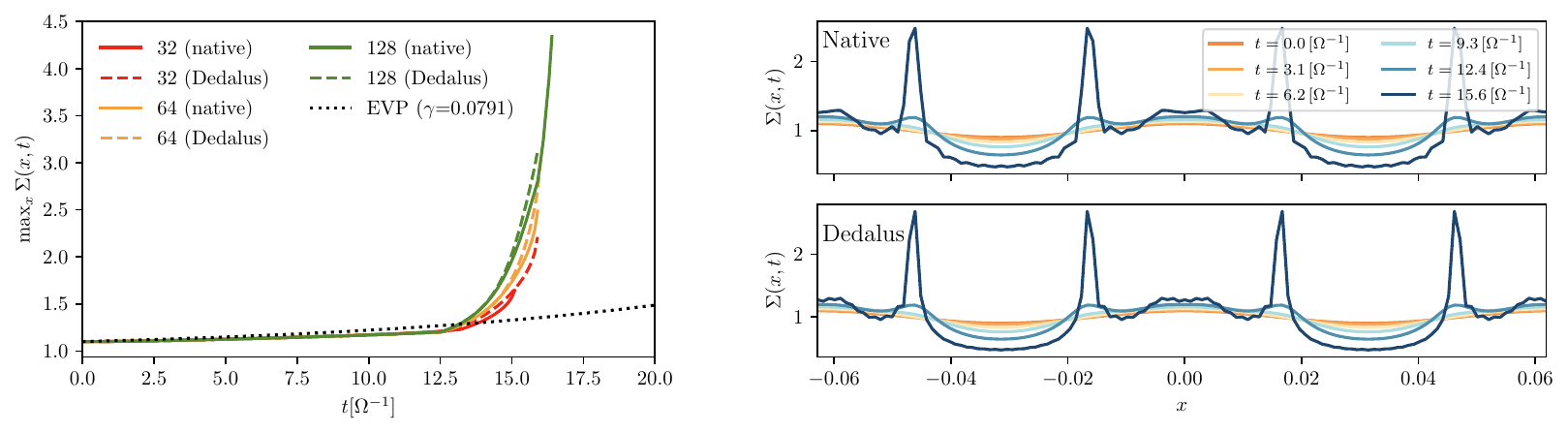}
    \caption{Nonlinear evolution and blow-up for the diffusive instability. Evolution of the maximum dust surface density for resolutions 32, 64, and 128 using both the native IMEX solver and \texttt{Dedalus} (left), and snapshots of $\Sigma(x,t)$ for the $N_x = 128$ runs at selected times for the native solver (upper right), and \texttt{Dedalus} (lower right). The initial $k=100$ eigenmode grows slightly below the linear rate (black dotted line), then transitions into a regime of super-exponential growth once nonlinear gradients dominate. The system does not saturate as steepening gradients reinforce the nonlinear dust-pressure term, while the viscosity is unable to regularize. Both solvers show the same qualitative evolution, where the initially sinusoidal mode sharpens into narrow spikes with width approaching the grid scale. Minor differences in peak sharpness reflect the different high-$k$ regularization properties of the two methods.}
    \label{fig:nonlinear_blowup}
\end{figure*}

Next, we investigate the diffusive instability's nonlinear evolution. For this purpose we again evolve the clean eigenmode solutions forward in time with $K = 100$, $\tau_\mathrm{s} = 0.1$, $\beta_\mathrm{visc} = \beta_\mathrm{diff} = -2.5, \delta = \alpha = 10^{-5}$, $\alpha_\mathrm{g} = 10^{-4}$, and $\varepsilon = 1$ with $L_x = 4\pi/k$. We test several resolutions.  Figure~\ref{fig:nonlinear_blowup} shows the evolution of the dust surface density amplitude (left panel) and the mode structure during selected snapshots (right panels) for both the native IMEX solver and \texttt{DEDALUS}. Eigenmodes are initialized at a perturbation amplitude of $\hat{\Sigma}_\mathrm{d} - \Sigma_\mathrm{d,0} = 0.1$, which initially grow slightly below the linear rate, followed by super-exponential growth once $\Sigma$ develops noticeable spatial gradients, culminating in nonlinear blowup.

The instability does not saturate as the dominant nonlinear term in Eq.~\eqref{eq:nonlinear_dust_vx} is $(1/\Sigma)\partial_x [(D^2/\Sigma)(\partial_x \Sigma)^2]$ scales as $D^2 (\partial_x\Sigma)^2/\Sigma^2$ plus higher derivatives. The adopted closure relation of $D \propto \Sigma^{\beta_\mathrm{diff}}$ with $\beta_\mathrm{diff} < 0$ (which drives the instability in the first place), has $D$ respond sharply to changes in $\Sigma$: In compressive regions $\partial_x \Sigma$ becomes large and negative, leading to strong pressure, that drives even stronger compression. The continuity equation then just transports $\Sigma$ into its gradients leading to finite-time collapse of $\Sigma$ into spikes, which can be seen in the right panels of Fig.~\ref{fig:nonlinear_blowup}. %This is analogous to Burgers-like shock steepening but with a sign that reinforces compression.
Also note that in our model the viscous terms that act on the velocity field are at most linear in spatial gradients. Thus, as density gradients steepen, the destabilizing $D^2 (\partial_x\Sigma)^2/\Sigma^2$ contribution overwhelms viscous diffusion. Therefore, dust viscosity, while linearly regularizing in the sense that it avoids growth on arbitrarily small scales (see Fig.~\ref{fig:instability_map}), cannot prevent nonlinear blow-up independent of the value of $\beta_\mathrm{visc}$.

\subsection{Growth of Random Noise}

\label{sect:randomnoise}

\begin{figure*}
    \includegraphics[width=\linewidth]{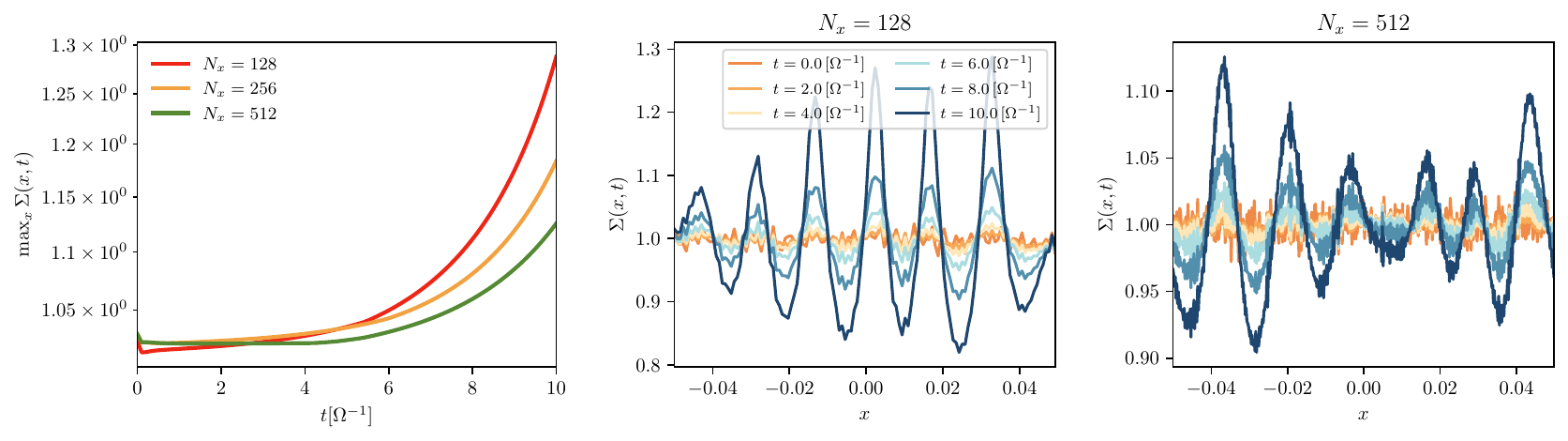}
    \caption{Evolution of a noise initial condition. Maximum dust density as a function of time (left) and snapshots of $\Sigma(x,t)$ for various times. Independently of resolution, the initial noise field organizes into a close-to pure Fourier mode with a wavenumber approaching the fastest-growing $k$ from linear theory.} 
    \label{fig:noise_evolution}
\end{figure*}

\begin{figure}
    \includegraphics[width=\linewidth]{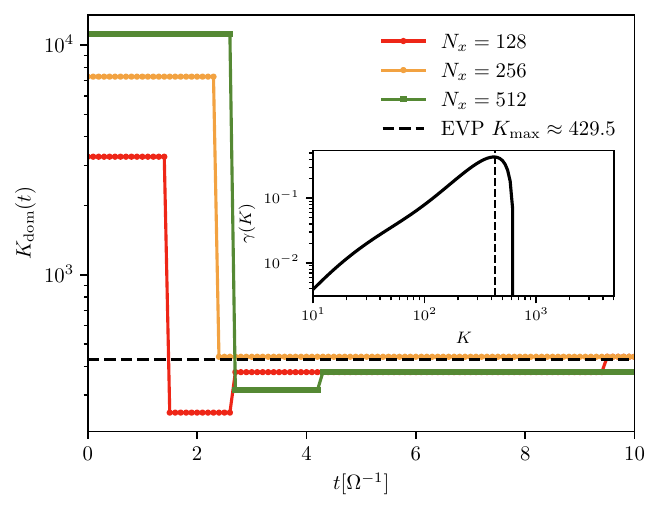}
    \caption{Selection of fastest-growing mode from the noise initial noise for the simulations shown in Fig.~\ref{fig:noise_evolution}. The dominant wavenumber approaches the fastest growing mode from linear theory. For reference, the inset shows the dispersion relation with that mode marked with a black dashed line.}
    \label{fig:noise_dominantk}
\end{figure}

For the final numerical test, we investigate the evolution of a random noise initial condition. This setup attempts to mimic the hypothesized application of the diffusive instability to the turbulent quasi steady state that characterizes the nonlinear outcome of the streaming instability within the boundaries of our model, i.e. the 1-dimensionality and lack of background drift. 

We initialize the $\Sigma$-field with Gaussian noise of initial amplitude $10^{-2}\Sigma_0$, which then interacts with the velocity fields initially set to $\hat{\bm{v}} = 0, u_y=0$. Parameters are set to $\tau_\mathrm{s} = 0.1, \beta_\mathrm{visc} = \beta_\mathrm{diff} = -2.5, \delta = \alpha = 10^{-5}$, $\alpha_\mathrm{g} = 10^{-4}$, and $\varepsilon = 1$, conducive to diffusive instability driven by the diffusion slope. We choose a domain size of $L_x = 0.1 H$, test three different resolutions $N_x \in \{128, 256, 512\}$, and stop the integration before nonlinear collapse.

Figure~\ref{fig:noise_evolution} shows that after an initial transient adjustment, the noise evolution is set by the linear instability, and the maximum surface density grows exponentially. The initial random perturbation contains a broad range of wavenumbers, however, over time, the fastest-growing Fourier mode of the instability emerges, independent of resolution. This is explicitly summarized in Fig.~\ref{fig:noise_dominantk}, where the evolution of the dominant wavenumber in the simulation is compared to the fastest-growing mode from the linear theory. Across resolutions, the instability robustly singles out this mode, confirming that the system behaves as a small-scale instability rather than noise-driven fragmentation process.

\section{Absence of Saturation Mechanisms}
\label{sect:absenceofsaturation}

The model as presented in Sect.~\ref{sect:model} leads to a linear instability with appreciable growth rates for plausible parameter choices. However, as discussed in Sect.~\ref{sect:nonlinaerblowup}, the instability does not saturate at high amplitudes but instead collapses into sharp spikes. Since this is a feature of the model equations rather than a numerical limitation, it stands to reason that the current model is inconsistent with simulations of nonlinear streaming instability. There, the close-to axisymmetric dust bands do, in fact, saturate and are not (generally) seen to collapse down to the resolution limit as would be predicted by the present model. 

We point out several key shortcomings when it comes to applying our model to the multidimensional streaming instability simulations. First, our model does not include a radial gas pressure gradient, as a background streaming motion is not required for the diffusive instability. On the other hand, in simulations of nonlinear streaming instability filaments, such a pressure gradient is generally included, because it sets the background equilibrium drift between gas and dust fluids which is a necessary condition for streaming instabilities. In the linear regime of the streaming instability, the perturbations do not modify the background drift, and the instability grows exponentially. However, in the nonlinear regime, dust pile-up induced back-reaction may accelerate the gas such that the background drift is appreciably reduced, weakening streaming instability growth rates where dust concentration is the highest. In contrast, our model does not possess such a self-regulating mechanism as the diffusion slope remains unmodified at high dust concentrations. Still, drift reduction alone is not sufficient to prevent runaway sharpening, as the linear streaming instability can still formally operate at filament edges. Multidimensional simulations show that parasitic instabilities also contribute to saturation \citep[see e.g.,][]{Johansen2007}. In particular, sharpening density contrasts leads to the development of velocity shear which in turn can drive Kelvin-Helmholtz Instabilities \citep[][]{Sekiya2000, Chiang2008, Gerbig2020}. These, as well as eddy diffusion in nonlinear streaming turbulence \citep[e.g.,][]{Schreiber2018, Klahr2020}, are not captured by our one-dimensional model. 

\subsection{A Saturating Diffusion Closure}

\begin{figure}
    \includegraphics[width=\linewidth]{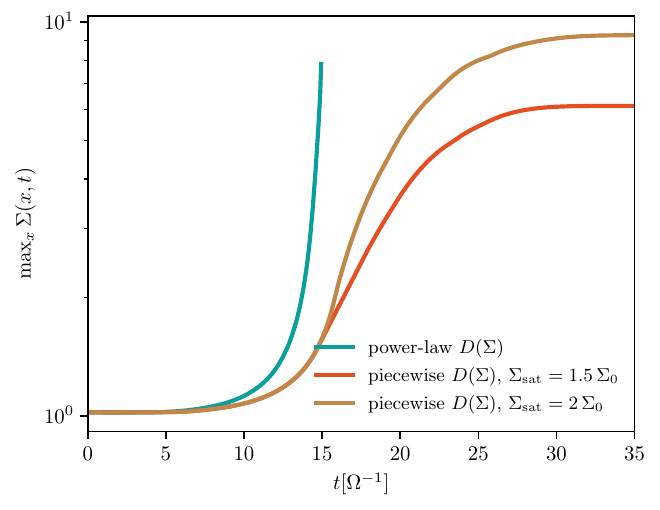}
    \caption{Evolution of the maximum surface density for three noise runs comparing the default power-law diffusion closure (teal, also see Fig.~\ref{fig:noise_evolution}), with piecewise diffusion closures that flatten at low and high density, see Eq.~\eqref{eq:piecewise_closure}. The runs are otherwise identical, and use $N_x = 512$. While the power-law model undergoes collapse, the piecewise closures eliminate blowup and lead to saturation.}
    \label{fig:piecewise_noise}
\end{figure}

The simplest way to modify the system's nonlinear behavior is to impose a closure relation that enforces saturation at high dust densities. Specifically, we assert a piecewise diffusion law $D(\Sigma) = D_0 f_\mathrm{pw}(\Sigma)$ and viscosity law $ \nu(\Sigma) = \nu_0 f_\mathrm{pw}(\Sigma)$ with
\begin{align}
\label{eq:piecewise_closure}
    f_\mathrm{pw}(\Sigma) = \begin{cases}
        1 & , \Sigma < \Sigma_\mathrm{0} \\
        \left(\Sigma/\Sigma_0\right)^{\beta_\mathrm{diff}} &, \Sigma_\mathrm{0} < \Sigma < \Sigma_\mathrm{sat} \\
        \left(\Sigma_\mathrm{sat}/\Sigma_0\right)^{\beta_\mathrm{diff}} & , \Sigma_\mathrm{sat} < \Sigma
    \end{cases}
\end{align}
so that the destabilizing negative slope operates only in a finite density interval and flattens at both low and high $\Sigma$.
While the high-$\Sigma$ saturation plateau is constructed ad-hoc, the low-$\Sigma$ plateau has some motivation in concrete diffusion measurements in streaming turbulence \citep[][]{Schreiber2018, Gerbig2023}. We emphasize that this prescription is not intended as a physical model of streaming turbulence per se, but rather as a demonstration that introducing a cutoff in the diffusion slope is sufficient to eliminate the finite-time blowup seen in the pure power-law model. For this purpose, we test two different values for $\Sigma_\mathrm{sat}/\Sigma_0 \in \left\{1.5, 2\right\}$ and evolve the noise initial condition at $N_x = 512$ in Fig.~\ref{fig:noise_evolution} forward in time. Fig.~\ref{fig:piecewise_noise} shows the evolution of the maximum surface density for those runs compared to the default power-law closure relation. With the piecewise diffusion closure, the instability no longer experiences collapse. Instead, it produces narrow filaments with densities controlled by the saturation scale $\Sigma_\mathrm{sat}$. In this sense, the saturation amplitude is externally imposed rather than a feature of the instability, underscoring that additional physics is required to predict filament widths and peak densities self-consistently. Indeed, the saturation density profiles remain sharper than in full streaming instability simulations, which may be attributed to the 1D geometry and the absence of parasitic instabilities as discussed above.

\section{Discussion}
\label{sect:discussion}

We revisited the diffusive instability proposed by \citet[][]{Gerbig_Lin_Lehmann_2024} using an updated, gas-inclusive formulation and nonlinear numerical experiments. Allowing an incompressible, viscous gas to respond azimuthally and including dust--gas feedback does not remove the instability: the basic criterion for diffusion-slope-driven growth remains $\beta_\mathrm{diff}\lesssim -2$ for $\tau_\mathrm{s}\ll 1$, and the overall structure of the unstable branches in $(\beta_\mathrm{diff},\tau_\mathrm{s})$--space, see Fig.~\ref{fig:instability_map}, is largely unchanged compared to the dust-only limit (see Fig.~\ref{fig:linearevolution_dustvgas}). Gas backreaction can modestly modify growth rates, with the largest effects occurring when the gas viscous response is weak and the dust-to-gas ratio is high. For parameter choices motivated by streaming-instability turbulence, however, the existence and character of the instability remain governed primarily by the effective density-dependence of dust transport coefficients.

Our nonlinear calculations highlight a key limitation of the present closure. With power-law transport coefficients $D,\nu\propto\Sigma^{\beta}$ and $\beta_\mathrm{diff}<-2$, the nonlinear evolution is generically prone to finite-time steepening and collapse into narrow spikes, see Fig.~\ref{fig:nonlinear_blowup}. This behavior %is not a numerical artifact: it appears in both our native IMEX solver and \texttt{Dedalus}, and
reflects the fact that the dominant nonlinear dust-pressure term strengthens as gradients sharpen, while viscous terms enter at lower order in spatial gradients and cannot provide a comparable nonlinear regularization. Because of this, the current 1D model does not qualitatively reproduce the nonlinear outcome of multidimensional streaming-instability simulations, where dense filaments saturate at finite amplitude.

The mismatch motivates two complementary interpretations. First, the diffusive instability may still be relevant as a linear organizing mechanism: our noise experiments show robust mode selection toward the fastest-growing wavenumber predicted by the eigenvalue problem. In this view, the instability could help set filament spacing during early organization of a turbulent dust field, while additional multidimensional physics determines the eventual filament thickness and peak density. Second, the closure itself may become invalid at high densities. In streaming turbulence, the effective diffusivity and viscous stress are not necessarily expected to continue decreasing indefinitely with increasing dust loading; instead, drift reduction, self-generated turbulence, and parasitic instabilities (e.g., Kelvin-Helmholtz instability) can act to limit further sharpening and enhance mixing. These mechanisms are absent by construction in our axisymmetric 1D framework, and their inclusion is a natural direction for future work.

As a minimal demonstration of how nonlinear saturation can be restored within the model, we introduced a piecewise closure in which the negative diffusion slope operates only over a finite interval $\Sigma_0<\Sigma<\Sigma_\mathrm{sat}$ and flattens at both low and high densities. This modification eliminates blowup, see Fig.~\ref{fig:piecewise_noise}, and produces filaments whose peak densities are controlled by the imposed $\Sigma_\mathrm{sat}$. While this confirms that a cutoff in the effective transport slope is sufficient to prevent collapse, it also emphasizes that the saturation amplitude is not predicted self-consistently. Connecting $\Sigma_\mathrm{sat}$ to measurable properties of streaming turbulence would be required to assess whether diffusive instabilities can quantitatively account for filament amplitudes and widths in simulations and, ultimately, for implications for planetesimal formation.

Overall, our results support the robustness of diffusive instabilities as a linear organizing mechanism in dusty protoplanetary disks whenever dust diffusion decreases sufficiently steeply with density. At the same time, we reiterate that realistic nonlinear outcomes depend sensitively on how turbulent diffusion behaves at high dust loading and on multidimensional saturation effects not captured by our model. Future work may therefore focus on (i) calibrating density-dependent transport closures directly from streaming instability simulations, and (ii) extending the model to include a background pressure gradient and incorporating multidimensional effects that, in the linear limit, allow a direct comparison to streaming instability, and enable parasitic instabilities and, possibly, self-consistent saturation.

%% Please use the acknowledgment and contribution environments. This will 
%% be anonomyized when the "anonymous" style option is used. 
\begin{acknowledgments}
This project is supported in part by Schmidt Sciences. \change{KG thanks Nikki Tebaldi for support with the code base.} \change{MKL is supported by the National Science and Technology Council (grants 114-2811-M-001-022-, 114-2112-M-001-018-, 
114-2124-M-002-003-, 115-2124-M-002-014-), an Academia 
Sinica Career Development Award (AS-CDA-110-M06), and an Academia Sinica Grand Challenge Seed Grant (AS-GCS-115-M02). } 
\end{acknowledgments}

\software{\change{\texttt{diff-inst} \citep{Gerbig_diffinst_zenodo}}, \texttt{Dedalus} \citep{Burns2020}, NumPy \citep{Harris2020}, Matplotlib \citep{Hunter2007}, CMasher \citep{vanderVelden2020}}.


%\appendix

%% For this sample we use BibTeX plus aasjournalv7.bst to generate the

\bibliography{references}{}
\bibliographystyle{aasjournalv7}


\end{CJK*}
\end{document}